\documentstyle[aasms4]{article}
\def\lesssim{\mathrel{\hbox{\rlap{\hbox{\lower4pt\hbox{$\sim$}}}\hbox{$<$}}}}
\def\gtrsim{\mathrel{\hbox{\rlap{\hbox{\lower4pt\hbox{$\sim$}}}\hbox{$>$}}}}
\slugcomment{to appear in {\it The Astrophysical Journal}, November 2000}
\begin{document}

\title{Formation of an Infalling Disklike Envelope
 around a Protostar}
\author{Fumitaka Nakamura}
\affil{Faculty of Education and Human Sciences, 
Niigata University, 8050 Ikarashi-2, Niigata, 950-2181, Japan}

\begin{abstract}
We examined the gravitational contraction of
 isothermal molecular cloud cores with slow rotation 
 by means of two-dimensional numerical simulations.
Applying a sink-cell method, 
 we followed the evolution of the cloud cores
 up to the stages at which most of the matter
 accretes onto the central region (i.e.,
 a protostar and a rotationally-supported circumstellar disk).
We show that both an infalling disklike envelope 
 and a rotationally-supported disk around the central star
 are natural outcome of the gravitational contraction of 
 a {\it prolate} cloud core with slow rotation.
The early evolution of the infalling envelopes 
 resembles sheet models recently proposed
 by Hartmann and coworkers.
In the infalling disklike envelope, the radial profiles of the 
 density, radial velocity, and azimuthal velocity can be 
 approximated by $\rho\propto r^{-1.5}$, $v_r\propto r^{-0.5}$,
 and $v_\varphi\propto r^{-1}$, respectively.
The fate of the infalling envelopes is also discussed.
\end{abstract}


\section{Introduction}

Recent high-resolution interferometric observations
 have suggested that disklike envelopes are common structures 
 around embedded young stars
 (e.g., Hayashi, Ohashi, \& Miyama 1993; Ohashi et al. 1996).
In the disklike envelopes, infall motions seem to 
 dominate over the rotation around the young stars, 
 indicating that the disklike envelopes
 are not rotationally-supported.
Thus, embedded young stars are most likely to be 
 undergoing dynamical accretion from their disklike envelopes
 onto quasi-static protostellar cores and
 rotationally-supported circumstellar disks.
The typical infalling disklike envelope 
 has a size of the order of $10^3$ AU, which
 is an order of magnitude larger in size than 
 a rotationally-supported circumstellar disk
 observed in the dust emission.
The infalling envelopes are expected
 to play a crucial role in the formation processes of stars 
 as well as their surrounding rotationally-supported disks.
It is thus important to understand the formation and evolution 
 of the infalling disklike envelopes.

Several authors have proposed the formation mechanisms of such 
 infalling disks around young stars
 by means of numerical hydrodynamics. 
Galli \& Shu (1993) examined the gravitational 
 contraction of magnetized spherical cloud cores 
 [see also Boss (1987) and Yorke \& Bodenheimer (1999)
 for the evolution of non-magnetized rotating spherical 
 cloud cores].
They showed that the magnetic field impedes the contraction 
 perpendicular to the field lines and thus 
 the disklike structures develop in the infalling envelopes.
In their model, the magnetic field is essential in the 
 formation of the infalling disks.
Hartmann et al. (1994) proposed another process for the formation 
 of the infalling disks.
They showed that even in the absence of 
 the magnetic field, initially sheetlike cloud cores 
 collapse to form highly disklike envelopes which
 have similar structures to Galli \& Shu (1993)'s disks.
Their calculations indicate that the cloud geometry plays an 
 important role in the formation of the infalling envelopes. 

Radio observations have shown that 
 most molecular cloud cores are elongated, implying 
 on statistical grounds that most of them are prolate
 (Myers et al. 1991).
Bonnell, Bate, \& Price (1996) thus examined
 the gravitational contraction 
 of non-rotating prolate cloud cores.
They found that in a prolate cloud core close to virial 
 equilibrium, the pressure force retards
 the contraction preferentially along its minor axis. 
As a result, an extended disklike structure,
 qualitatively similar to
 Hartmann et al. (1994)'s sheet model, is formed.

Although these calculations have demonstrated
 the formation processes
 of the infalling disks in molecular cloud cores, 
 it is still not clear how the infalling disks form and evolve 
 subsequently and 
 how rotationally-supported circumstellar disks form
 in the infalling disklike envelopes.

In this paper, 
 we examine the gravitational collapse of prolate
 cloud cores with slow rotation.  
We show that both the infalling disklike envelope 
 and rotationally-supported circumstellar disk
 are natural outcome of the gravitational contraction of 
 a {\it prolate} cloud core.
In \S \ref{sec:model}, we describe our model and numerical method.
Numerical results are given in \S \ref{sec:results}.
In \S \ref{sec:mass-infall rate}, we discuss the evolution of 
the mass infall rate in the disklike envelope.
Finally we discuss implications to observations.

\section{Models and Numerical Methods}
\label{sec:model}

Radio observations have revealed that
 most molecular clouds have more or less
 filamentary structures. 
Recent numerical simulations have also shown that
 such filaments are ubiquitous features in
 self-gravitating molecular clouds
 (e.g., Klessen, Burkert, \& Bate 1998).
According to a linear stability analysis,
 the linear perturbation grows fastest in time 
 when its longitudinal wavelength is several times
 as long as the effective cloud diameter.
It is worth noting that in the absence of the 
 strong magnetic fields and fast rotation,
 the filamentary cloud is unstable to 
 linear perturbations whose wavelengths are more than twice
 the effective cloud diameter
 (Matsumoto, Nakamura, \& Hanawa 1994).
Hence, the fragments tend to become prolate
 at the epoch of the fragmentation.
We investigate the evolution of such prolate fragments 
 formed through the fragmentation of filamentary clouds.

The model cloud is the same as that of
 Matsumoto, Hanawa, \& Nakamura (1997).
We consider an isothermal cylindrical cloud in which 
 the density is uniform along the cylinder axis.
We assume that the cloud is in hydrostatic equilibrium
 in the radial direction and that the cloud 
 is rotating around the cylinder axis.
The ratio of the centrifugal force to the pressure force,
 $\beta$, is assumed to be spatially uniform.
The radial distributions of the density and velocity are
 then given by 
\begin{equation}
\rho =\rho _0 \left(1+\frac{r^2}{8H^2}\right)^{-2} \; ,
\end{equation}
and 
\begin{equation}
\mbox{\boldmath$v$} = (v_r, v_\varphi, v_z)=
 \left[0, r\Omega _0\left(1+\frac{r^2}{8H^2}\right)^{-1/2},
 0\right] \; ,
\end{equation}
 respectively, where $H^2=(1+\beta)c_s^2/(4\pi G\rho _0)$,
 $\Omega _0^2=2\pi G\rho _0 \beta /(1+\beta )$,
 $c_s$ is the isothermal sound speed,
 $G$ is the gravitational constant, 
 and $\rho _0$ is the initial central density.
It should be noted that in this model, most of the matter
 rotates nearly uniformly ($v_\varphi \propto r$) and
 only the outer low-density region with a small mass
 rotates differentially ($v_\varphi = {\rm const.}$).

The initial model of our simulation is obtained by adding 
 a velocity perturbation onto the equilibrium model. 
The velocity perturbation changes sinusoidally 
 with the wavelength of $\lambda _z$, 
\begin{equation}
\delta v_z=v_0 \sin (2\pi z/\lambda _z) \; ,
\end{equation}
 where $v_0$ is the amplitude of the perturbation and
 is set to 5\% of the sound speed.
In the following, we take $\rho _0$ and $c_s$ as the units of 
 density and velocity, respectively.
Then, the initial model is specified by the two parameters of 
 $\beta$ and $\lambda _z$.
For the typical models, 
 the wavelength $\lambda _z$ is taken to be
 15.5 $c_s/\sqrt{2\pi G\rho_0}$,
 which is almost equal to those of
 the fastest-growing linear modes
 for non-rotating or slowly-rotating clouds
 (Matsumoto et al. 1994).
Observationally, most molecular cloud cores
 are likely to be rotating quite slowly (Goodman et al. 1993). 
Thus, we adopt the small values of $\beta$ ($=0-0.1$).
In Table \ref{tab:model}, we summarize
 the initial conditions of the models calculated in this paper.

We solved numerically the time evolution of the model cloud
 with the spatially and temporally
 second-order-accurate hydrodynamical
 code described by Matsumoto et al. (1997).
Since our model keeps mirror symmetry at $z=0$ and $\lambda _z/2$
throughout the evolution, we restricted the computational
 domain  in the interval of $0 \le z \le \lambda _z/2$.  
We set the fixed boundary conditions
 at $r=0, \ R_{\rm max}$ and
 $z=0, \ Z_{\rm max}$, where $R_{\rm max}$
 and $Z_{\rm max}$ denote
 the lengths of the computational box in the $r$- and
 $z$-directions, respectively.
For all the models, 
 we take $R_{\rm max}=2Z_{\rm max}=\lambda _z$.
The effect of the fixed boundary is very small.
We have calculated the evolution of the models
 with larger $R_{\rm max}$ and have confirmed
 that the numerical results are not changed.

Many numerical simulations have shown that
 the gravitational collapse of an isothermal cloud accelerates
 in the central high-density region.  
Thus, the characteristic length and time scales
 shorten as the collapse proceeds,
 this phase being referred to as runaway collapse phase.  
Therefore, it is difficult to resolve the central high-density
 region sufficiently at the late stages of 
 the computation even if the cloud evolution is
 followed with very fine grids.  
In real evolution, as the collapse progresses, 
a protostellar core is formed at the center
 and thereafter the envelope gas begins to accrete
 onto the central core, 
 this phase being referred to as the accretion phase.  

To pursue the evolution of the infalling envelope gas,
 we adopted a sink-cell method 
 when the central density reaches a reference density
 $\rho_{\rm sink}$ [see, e.g., Boss \& Black (1982) and Tomisaka (1996)].  
In this paper, we take $\rho _{\rm sink}=10^4\rho _0$ and 
 the innermost region of $0\le r\le 5\Delta r$
 and $0\le z\le 5\Delta z$ is assigned to the sink cells, where 
 $\Delta r$ and $\Delta z$ denote the grid spacings in the $r$-
 and $z$-directions, respectively.
The excess density ($\rho -\rho_{\rm sink}$) and
 the corresponding momentum are removed from 
 the sink cells in every time step.  
The removed mass is added to a central point mass that  
 affects the matter in the envelope only through gravitation.

As mentioned above, we take $\rho _0 =1$, $c_s =1$,
 and $2\pi G=1$.
Then, the units of time, length, and mass are given as
\begin{eqnarray}
t_{\rm u}&=&(2\pi G\rho _0)^{-0.5} \nonumber \\
&=&2.5\times 10^4 {\rm yr} \left(\frac{n _0}
 {10^6{\rm cm^{-3}}} \right)^{-0.5} \, , \\
r_{\rm u}&=&c_s(2\pi G\rho _0)^{-0.5} \nonumber \\
&=&1.1\times 10^3 {\rm AU}
 \left(\frac{c_s}{0.2{\rm km \, s^{-1}}}\right)
 \left(\frac{n _0}
 {10^6{\rm cm^{-3}}}
 \right)^{-0.5} \, , \\
m_{\rm u}&=&\rho _cr_{\rm u}^3 \nonumber \\
&=&7.6\times 10^{-3} M_\odot
 \left(\frac{c_s}{0.2 \, {\rm km \, s^{-1}}}\right)^3
 \left(\frac{n _0}
 {10^6{\rm cm^{-3}}} \right)^{-0.5} \, , 
\end{eqnarray}
 respectively, where 
 $n_0\equiv \rho _0/(\mu m_{\rm H})$ is the number density,
 $\mu$ is the mean molecular weight and taken to be 2.3,
 $m_{\rm H}$ is the mass of a hydrogen atom.

\section{Numerical Results}
\label{sec:results}

\subsection{Prolate Cloud Cores without Rotation}
\label{subsec:norotation}

In this subsection, we explore the evolution of
 non-rotating models.
In the following, we mainly show the numerical results of 
 model B as a typical example.
In this model, the wavelength $\lambda _z$ is almost equal to 
 that of the fastest growing linear perturbation
 (e.g., Nakamura, Hanawa, \& Nakano 1993).
This model is essentially the same as 
 model D of Nakamura, Hanawa, \& Nakano (1995) 
 who followed the cloud contraction during
 the runaway collapse phase.
We extend their calculations to considerably 
 later stages, i.e., the accretion phase, 
 by applying the sink-cell method.
We pursued the accretion phase evolution until most 
 of the envelope gas accretes onto the central point mass.

\begin{figure}
\epsscale{0.6}
\plotone{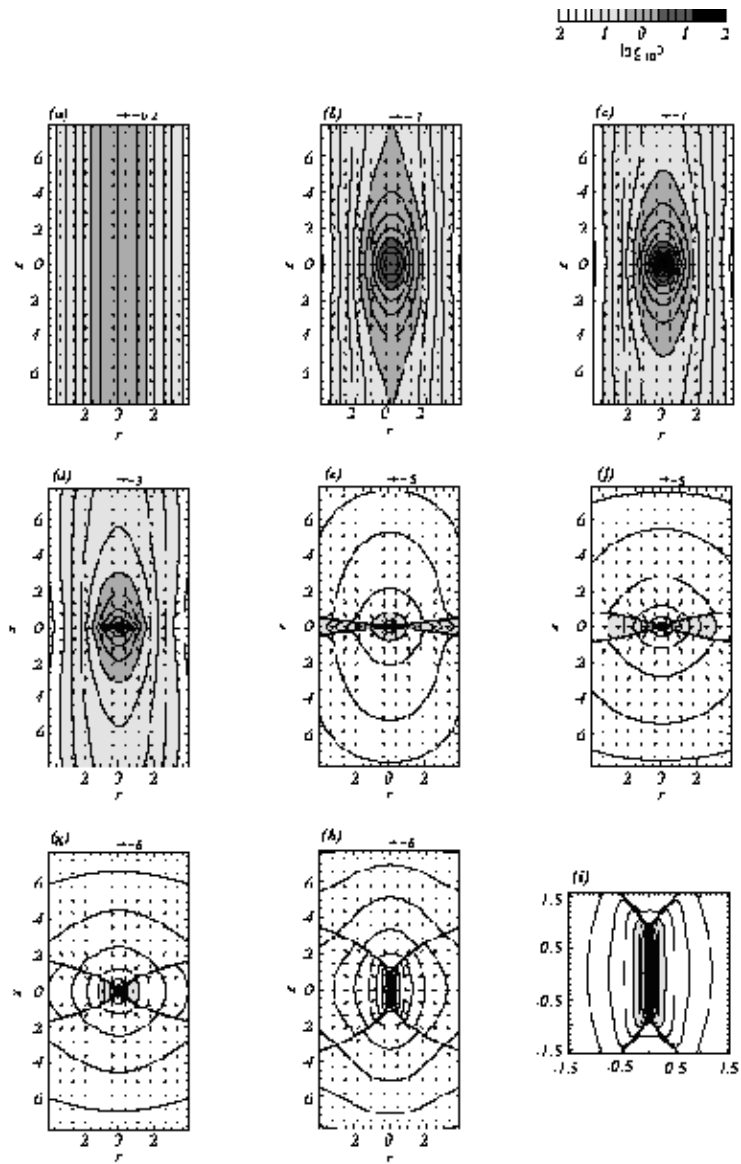}
\caption{Density and velocity distributions in the $r-z$ plane
 for model B at eight different stages.
This model has the initial parameters of $\beta=0$ and 
$\lambda _z=15.5r_{\rm u}$.
The gray scale indicates the density distribution
 in a logarithmic scale.
The arrows show the velocity distributions which are normalized
 separately in each panel.
Panels (a) through (h) show the cross sections at 
 (a) $t-t_{\rm core}=-7.23 t_{\rm u}$
 ($\rho_c=\rho_0$), (b) $-0.59$ ($\rho_c=10\rho_0$), 
 (c) 0.00 ($\rho_c=10^4\rho_0$),
 (d) 0.93 ($M_0\simeq 0.3 M_{\rm t}$),
 (e) 2.25 ($M_0\simeq 0.5 M_{\rm t}$), 
 (f) 4.49 ($M_0\simeq 0.7 M_{\rm t}$),
 (g) 8.28 ($M_0\simeq 0.86 M_{\rm t}$), and 
 (h) 10.88 ($M_0\simeq 0.9 M_{\rm t}$), respectively, 
 where $t-t_{\rm core}$ is the time measured from the epoch of core
 formation, $\rho _c$ is the central density, and $M_{\rm 0}$ is 
 the mass of the central point-like object.  
The first three panels represent the cross sections
 during the runaway collapse phase,
 while the others are during the accretion phase.
Panel (i) denotes the enlargement of the central region 
 of panel (h).
If we adopt $\rho _0=10^6$cm$^{-3}$ and $c_s=0.2$km s$^{-1}$, 
 then the units of time, length, and mass are given as
 $t_{\rm u}=(2\pi G\rho_ 0)^{-0.5}=2.5\times 10^4$ yr, 
 $r_{\rm u}=c_s t_{\rm u}=1.1\times 10^3$ AU, and 
 $m_{\rm u}=\rho _0 r_{\rm u}^3=7.6\times 10^{-3}M_\odot$, 
 respectively.
The total cloud mass is equal to
 $1.9\times 10^2m_{\rm u}=1.5M_\odot$.
\label{fig:cross sections1}
}
\end{figure}

Figure \ref{fig:cross sections1}
 shows the density and velocity distributions
 in the $r-z$ plane at eight different stages.
The first three panels show the cross sections of the cloud
 during the runaway collapse phase,
 while the others show those during the accretion phase.
At the early stages, the cloud collapses preferentially
towards the $z=0$ plane
 (Figs. \ref{fig:cross sections1}b and \ref{fig:cross sections1}c). 
As the collapse proceeds, the central 
 high-density region becomes spherical, whereas
 the low-density region remains prolate.
The central spherical region continues to contract
 towards the center.
Such evolution is the same as that of model D by
 Nakamura et al. (1995).

When the central density reaches $10^4\rho_0$, 
 the sink cell method is applied and the central point-like
 object forms.
The mass of the central point-like object monotonously
 increases with time.
Since the vertical flow is dominant near the center, 
 the disklike envelope is formed around the central
 point-like object and grows in radial extent
 (Figs. \ref{fig:cross sections1}d and \ref{fig:cross sections1}e).
A shock wave is also formed at the disk surface.
By the stage at which the mass of the central point-like object
 reaches about $0.7 M_{\rm t}$,
 the infalling disk extends 
 radially up to $r \sim 4-5r_{\rm u}$, where $M_{\rm t}$ denotes 
 the total cloud mass.
In the disk, the gravitational force dominates the pressure force 
 in the radial direction, while 
 in the vertical direction, the pressure force 
 becomes comparable to the gravitational force.  
Therefore, the gas flow progresses mainly
 in the  radial direction inside the disk.
Such an non-equilibrium disk resembles pseudo-disks
 found by Galli \& Shu (1993), Hartmann et al. (1994),
 and Bonnell et al. (1996).

After the mass of the central point-like object reaches 
 about $0.7 M_{\rm t}$,
 the radius of the infalling disk becomes 
 shorter because of dominant radial infall 
 (Figs. \ref{fig:cross sections1}f and \ref{fig:cross sections1}g).
The shock front lifts up vertically because of 
 dominant pressure force in the vertical direction.
Thereafter, the shock front 
 reaches the $z$-axis and thus
 the matter in the infalling disk collapses to form 
 a thin needle-like (prolate) object along the $z$-axis
 (Figs. \ref{fig:cross sections1}h
 and \ref{fig:cross sections1}i).

We also followed the evolution of models A and C
 to examine the effects of $\lambda$
 on the formation and evolution of the disklike envelope.
It is found that the evolution is qualitatively
 similar to that of model B.
In both models, the disklike envelopes are formed and then 
 they evolve into thin needle-like objects.
For the model with larger $\lambda$, the vertical infall
 is more dominant and thus the disk is more extended
 in the radial direction.

\subsection{Prolate Cloud Cores with Slow Rotation}
\label{subsec:rotation}

In this subsection, we explore the evolution of 
 slowly-rotating clouds.
In the following, we mainly show the numerical results of 
 model E as a typical example. 
In this model, the wavelength $\lambda _z$ is almost equal to 
 that of the fastest growing linear perturbation.
As shown below, the evolution of model E is qualitatively similar to
 that of model B except near the center,
 i.e., the cloud collapses to form an infalling disk in which 
 the gas flows mainly in the radial direction.

\begin{figure}
\epsscale{0.6}
\plotone{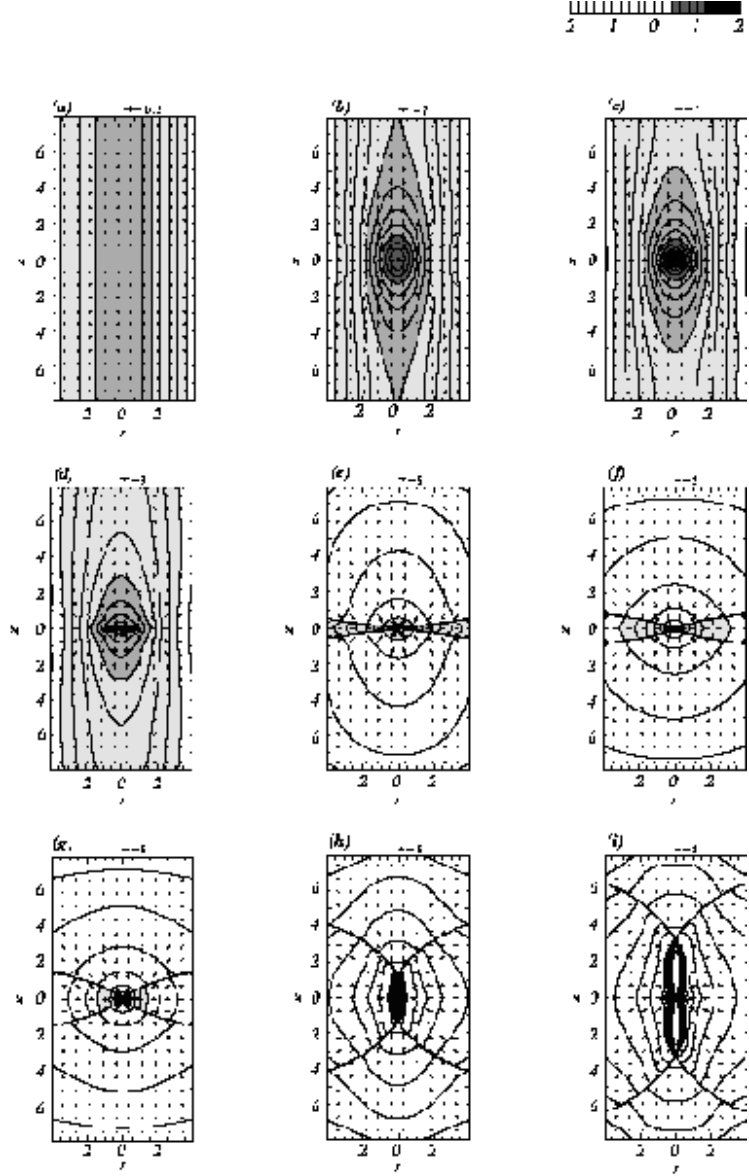}
\caption{Same as Figure \ref{fig:cross sections1} but for model E.
This model has the initial parameters of $\beta=0.05$ and 
$\lambda _z=15.5r_{\rm u}$.
Panels (a) through (i) show the cross sections at 
 (a) $t-t_{\rm core}=-7.26t_{\rm u}$
 ($\rho_c=\rho_0$), (b) $-0.59$ ($\rho_c=10\rho_0$), 
 (c) 0.00 ($\rho_c=10^4\rho_0$),
 (d) 0.98 ($M_0\simeq 0.3 M_{\rm t}$),
 (e) 4.71 ($M_0\simeq 0.7 M_{\rm t}$),
 (f) 6.59 ($M_0\simeq 0.79 M_{\rm t}$), 
 (g) 8.25 ($M_0\simeq 0.82 M_{\rm t}$), 
 (h) 11.74 ($M_0\simeq 0.84 M_{\rm t}$),and
 (i) 13.44 ($M_0\simeq 0.85 M_{\rm t}$), respectively, 
 where $t-t_{\rm core}$ is the time measured from the epoch of core
 formation, $\rho _c$ is the central density, and $M_{\rm 0}$ is 
 the mass of the central point-like object.  
If we adopt $\rho _0=10^6$cm$^{-3}$ and $c_s=0.2$km s$^{-1}$, 
 then the units of time, length, and mass are given as
 $t_{\rm u}=(2\pi G\rho_ 0)^{-0.5}=2.5\times 10^4$ yr, 
 $r_{\rm u}=c_s t_{\rm u}=1.1\times 10^3$ AU, and 
 $m_{\rm u}=\rho _0 r_{\rm u}^3=7.6\times 10^{-3}M_\odot$, 
 respectively.
The total cloud mass is equal to
 $2.1\times 10^2m_{\rm u}=1.6M_\odot$.
\label{fig:cross sections2}
}
\end{figure}

Figure \ref{fig:cross sections2} is the same as 
 Figure \ref{fig:cross sections1} but for model E.
During the runaway collapse phase, 
 the central high-density region becomes slightly oblate
 because of rotation, whereas
 the low-density region remains prolate.
In the central region, the rotation is not important
 in the cloud support.
Therefore, the central region continues to contract.

After the central core is formed, 
 the disklike envelope is formed and then expands in radius
 (Figs. \ref{fig:cross sections2}d and \ref{fig:cross sections2}e).
A shock wave is also formed at the disk surface
 and propagates towards the $z$-axis.
Such evolution is essentially the same as that of model B.
In the disk, the rotation is not important and therefore
 the gas flow progresses primarily in the radial direction.
As the contraction proceeds, 
 the centrifugal force becomes important at the innermost region 
 and a rotationally-supported disk is formed there.

\begin{figure}
\epsscale{1.0}
\plotone{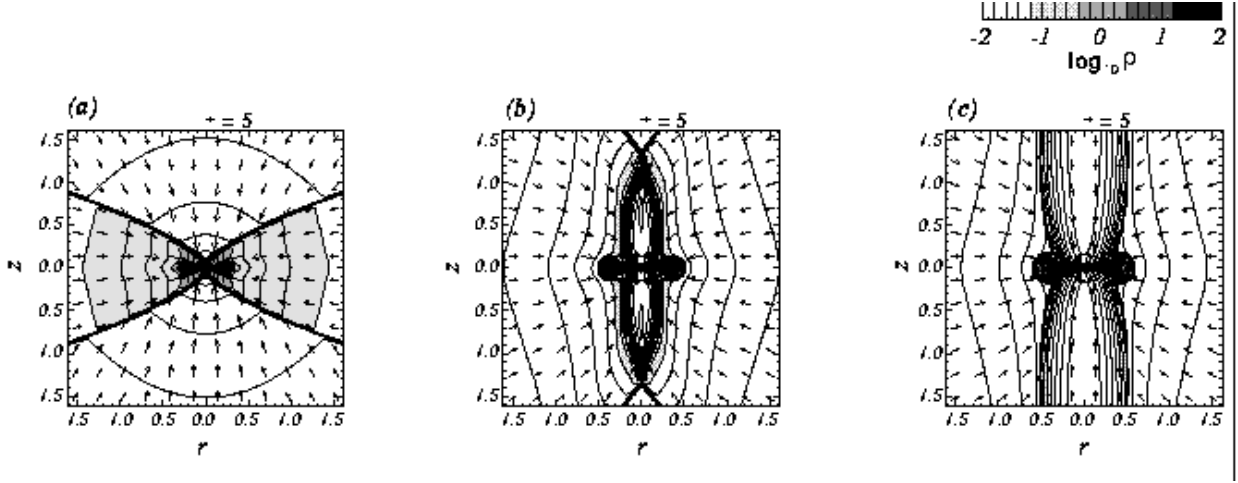}
\caption{Density and velocity distributions of the central region for
 model E. Panels (a), (b), and (c) denote the cross sections at 
 the same stages as panels (g), (h), and (i) of
 Figure \ref{fig:cross sections2}, respectively.
[(a) $t-t_{\rm core}=8.25t_{\rm u}$ ($M_0\simeq 0.82 M_{\rm t}$), 
 (b) 11.74 ($M_0\simeq 0.84 M_{\rm t}$), and
 (c) 13.44 ($M_0\simeq 0.85 M_{\rm t}$), respectively, 
 where $t-t_{\rm core}$ is the time measured from the epoch of core
 formation and $M_{\rm 0}$ is the mass of the central point-like object.]
\label{fig:cross sections3}
}
\end{figure}

After the mass of the central region reaches
 about $0.7M_{\rm t}$,
 the radius of the infalling disk becomes 
 shorter owing to dominant radial infall 
 (Figs. \ref{fig:cross sections2}f and \ref{fig:cross sections2}g).
Then, the matter in the infalling disk collapses to form 
 a funnel in which 
 the centrifugal force balances with the gravitational force 
 in the radial direction (Figs. \ref{fig:cross sections2}h and 
 \ref{fig:cross sections2}i). 
In Figure \ref{fig:cross sections3}, we show the enlargements of 
 the central region at the same stages as the last three panels of 
 Figure \ref{fig:cross sections2}.
A new shock front is also formed there
 and propagates in the radial direction
 as the gas with higher specific angular momentum accretes.
Such a structure resembles a bipolar cavity which 
 is often thought to be formed by the outflow from the central 
 young star.
This structure might play a significant role in the guidance
 of outflow from the central star at the late stages of the evolution.

The rotationally-supported disk is encased
 in two accretion shock fronts, both of which are 
 located several scale heights above the equatorial plane
 (Figs. \ref{fig:cross sections3}b and \ref{fig:cross sections3}c).
Such structure seems to be similar to that of Yorke \& Bodenheimer (1999).
Shock formation is explained as follows.
In the infalling disk, the accretion flow is
 dominant in the radial direction.
The radial infall stops at the funnel owing to 
 the centrifugal barrier and thus the outer shock front is formed.
Thereafter, the matter in the rotationally-supported disk
 collapses in the vertical direction
 to achieve hydrostatic equilibrium.
This infall produces the inner shock front.
In the innermost disk, angular momentum transfer is expected to
 be important in the accretion flow onto the central core, 
 although it is not included in the present calculations 
 (see Yorke \& Bodenheimer 1999).

We also pursued the evolution of models D and F
 to examine the effects of $\beta$.
It is found that the evolution is qualitatively
 similar to that of model E, although the size of the 
 rotationally-supported disk depends on $\beta$, i.e., 
 when $\beta$ is larger,
 the rotatinally-supported disk is wider.

\subsection{Effects of Cloud Geometry
 (Prolate vs. Oblate Cloud Cores)}

As shown in \S \ref{subsec:norotation} and \S \ref{subsec:rotation},
 a prolate cloud core collapses to form an infalling disk
 in which the accreting gas flows mainly in the radial direction.
In this subsection, to clarify the physics of the formation process of the
 infalling disk, we calculate the evolution of two representative
 models: (1) a prolate cloud and (2) an oblate cloud.
In the following, we consider non-rotating
 prolate and oblate clouds
 which have Gaussian-like density profiles as 
\begin{equation}
 \rho =\rho _{0} \exp \left[-\left(\frac{r^2}{R_0^2}+
 \frac{z^2}{Z_0^2}\right)\right] \; .
\end{equation}
At the initial state, the cloud is assumed to be static for simplicity.
This model has the two parameters of $R_0$ and $Z_0$.

\begin{figure}
\epsscale{1.0}
\plotone{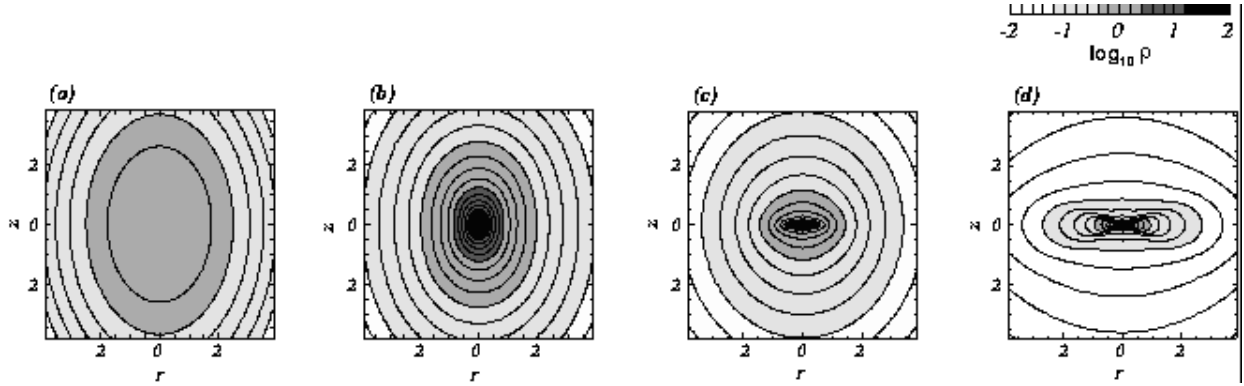}
\caption{Density and velocity distributions in the $r-z$ plane
 for the prolate cloud at four different stages. 
This model has the initial parameters of 
 $(R_0, Z_0)=(1.66r_{\rm u}, 2.49r_{\rm u})$.
The gray scale indicates the density distribution
 in a logarithmic scale.
Panels (a) through (d) show the cross sections at 
 (a) $t-t_{\rm core}=-2.06t_{\rm u}$ ($\rho_c=\rho_0$),
 (b) 0.00 ($\rho_c=10^4\rho_0$), 
 (c) 1.09 ($M_0\simeq 0.5 M_{\rm t}$),
 (d) 2.64 ($M_0\simeq 0.7 M_{\rm t}$), respectively,  
 where $t-t_{\rm core}$ is the time measured from the epoch of core
 formation, $\rho _c$ is the central density, and $M_{\rm 0}$ is 
 the mass of the central point-like object.  
\label{fig:prolate}
}
\end{figure}

\begin{figure}
\epsscale{1.0}
\plotone{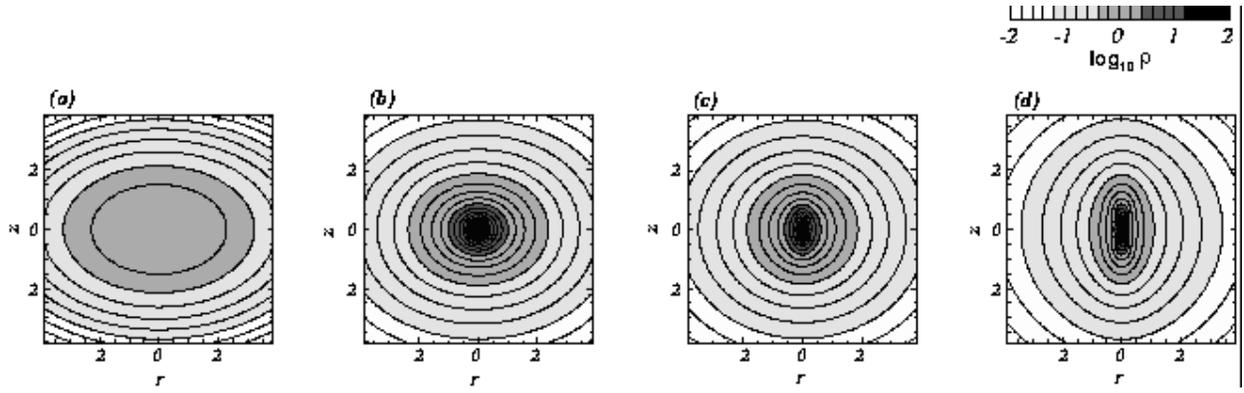}
\caption{Same as Figure \ref{fig:prolate} but for the 
 oblate cloud.
This model has the initial parameters of 
 $(R_0, Z_0)=(2.40r_{\rm u}, 1.60r_{\rm u})$.
The gray scale indicates the density distribution
 in a logarithmic scale.
Panels (a) through (d) show the cross sections at 
 (a) $t-t_{\rm core}=-2.08t_{\rm u}$ ($\rho_c=\rho_0$),
 (b) 0.00 ($\rho_c=10^4\rho_0$), 
 (c) 0.49 ($M_0\simeq 0.3 M_{\rm t}$),
 (d) 1.12 ($M_0\simeq 0.5 M_{\rm t}$), respectively, 
 where $t-t_{\rm core}$ is the time measured from the epoch of core
 formation, $\rho _c$ is the central density, and $M_{\rm 0}$ is 
 the mass of the central point-like object.  
\label{fig:oblate}
}
\end{figure}

Figures \ref{fig:prolate} and \ref{fig:oblate} show
 the evolutions of the density distributions in the $r-z$ plane for 
 the prolate and oblate clouds, respectively.
The prolate and oblate clouds have the model parameters of 
 $(R_0, Z_0)=(1.66r_{\rm u}, 2.49r_{\rm u})$ and 
 $(2.40r_{\rm u}, 1.60r_{\rm u})$, respectively.
In both models, the ratio of the major to minor axes
 is set to 1.5 and the ratio of the thermal energy
 to the gravitational energy is equal to $\alpha \sim 0.33$.
The total cloud masses are equal to
 $M_{\rm t}=1.44\times 10^2 m_{\rm u}$ 
 and $1.40\times 10^2 m_{\rm u}$ 
 for the prolate and oblate clouds, respectively.
The grid spacing is set to $\Delta r=\Delta z=0.02r_u$.
The number of grid points is taken to be $N_r \times N_z=512\times 1024$
In Figures \ref{fig:prolate} and \ref{fig:oblate},
 the first two panels show the cross sections 
 during the runaway collapse phase, while the rest are
 during the accretion phase.
In both models, the contraction progresses preferentially
 along the major axis.
Therefore, during the accretion phase,
 the initially prolate cloud collapses into an oblate core, 
 whereas the initially oblate cloud collapses into a prolate core.

This evolution is explained as follows.
At the initial state, the pressure force is less effective
 along the major axis because of shallower density gradient.
In other words, the net attractive force is stronger
 along the major axis.
Therefore, the cloud collapses preferentially
 along the major axis.

We also followed the evolution of the models with different
 axis ratios and $\alpha$.
It is found that as long as the cloud is close
 to virial equilibrium, the evolution is qualitatively similar 
 to those shown here (see also Bonnell et al. 1996).

As mentioned in \S \ref{sec:model},
 the fragments formed from filamentary clouds tend to become
 prolate at the epoch of fragmentation.
Therefore, as shown in \S \ref{subsec:norotation} and 
 \S \ref{subsec:rotation}, such a prolate fragment
 collapses to form an oblate core
 which thereafter evolves into a prolate core.
This evolution is consistent with those of Figures
 \ref{fig:prolate} and  \ref{fig:oblate}.

\section{Time-Dependent Mass Infall Rate}
\label{sec:mass-infall rate}

In this section, we discuss the evolution of 
the mass infall rate onto the center.

\begin{figure}
\epsscale{0.55}
\plotone{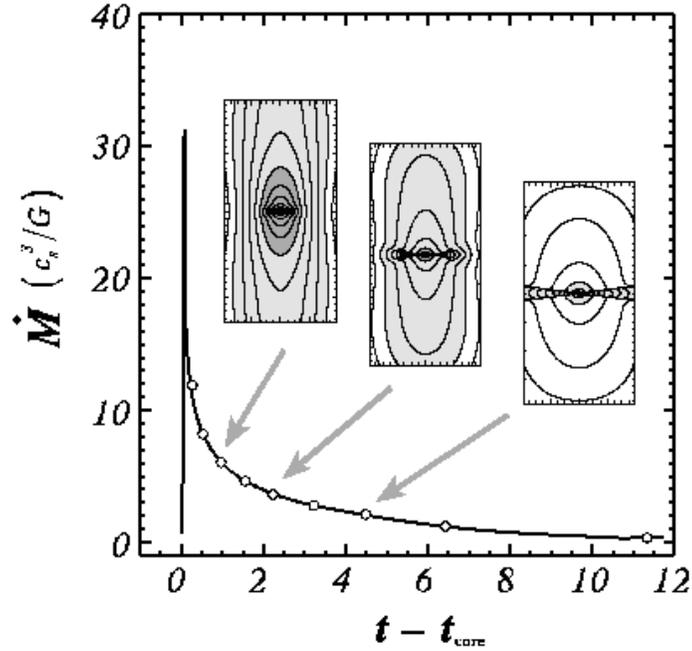}
\caption{
Mass infall rate of model B 
 as a function of time measured from the 
 epoch of core formation ($t-t_{\rm core}$).
If we adopt $\rho _0=10^6$cm$^{-3}$ and $c_s=0.2$km s$^{-1}$, 
 then the units of time and mass infall rate are given as
 $t_{\rm u}=(2\pi G\rho_ 0)^{-0.5}=2.5\times 10^4$ yr and  
 $c_s^3/G=1.9\times 10^{-6} M_\odot$ yr$^{-1}$, respectively.
Open circles denote the stages at which 
 $M_0 =0.1 M_{\rm t}$, $0.2M_{\rm t}$, 
 $\cdots$, $0.7M_{\rm t}$.
For comparison, we show the cross sections of the cloud 
 at the stages of $M_0 =0.3 M_{\rm t}$,
 0.5 $M_{\rm t}$, and 0.7 $M_{\rm t}$.
}
\label{fig:accretion}
\end{figure}

Figure \ref{fig:accretion} shows
 the evolution of the mass infall rate
 for model B.
The mass infall rate is time-dependent, in contrast to
 the prediction of similarity solutions
 (e.g., Shu 1977; Whitworth \& Summers 1985).
The mass infall rate takes its maximum at the very early stage 
 and then monotonously decreases with time.
This behavior is basically the same as that of
 Ogino, Tomisaka, Nakamura (1999)'s model
 with a small $\alpha$, where $\alpha$ is the ratio of 
 the gravitational force to the pressure force.
Such evolution is explained as follows.
During the runaway collapse phase, the central region tends to
 approach the Larson-Penston similarity solution
 (Larson 1969; Penston 1969)
 whose mass infall rate is temporally constant
 at $\dot{M} _{\rm LP}\sim 47 c_s^3/G$.
However, only the central small region converges
 onto the Larson-Penston solution by the epoch of core formation.
In the outer infalling envelope, the pressure effect is
 important and the infall is appreciably retarded.
Thus, after the central small region accretes onto
 the central point mass, the mass infall rate declines in time.

Observed young stellar objects (YSOs) are often classified 
 into several empirical evolutional stages from star-less
 dense cores to main-sequence stars
 (e.g., Andr\'e, Ward-Thompson, \& Barsony 1993).
The youngest observed YSOs are called Class 0 sources which
 are interpreted to be in a phase at which the envelope gas
 still has more mass than the central hydrostatic protostellar
 core.
More evolved YSOs are called Class I sources.
Recently, Bontemps et al. (1996) 
 suggested that if the CO outflow rates are proportional to 
 the mass infall rates onto the embedded young stars,
 the mass infall rates of Class 0 sources are a factor of 
 10 larger on average than those of Class I sources.
Applying a simplified pressure-free model, Henriksen, Andr\'e, \& Bontemps (1997)
 proposed that the observed dispersion of the mass infall rates
 for Class 0 sources is due to the time evolution of 
 the mass infall rates. 
In our isothermal model,
 the evolution of the mass infall rate is
 qualitatively similar to that of Henriksen et al. (1997), 
 although in our hydrodynamical model, 
 the pressure force significantly
 retards the rapid change of the mass infall rate.
As discussed by Ogino et al. (1999), the evolution of 
 the mass infall rate also depends on the initial conditions 
 of the clouds.
We will discuss those effects in a separate paper.

\section{Implications to Observations}

Recent observations
 have revealed the detailed structures of the infalling 
 envelopes around several embedded young stars 
 (e.g., Hayashi et al. 1993; Ohashi et al. 1996;
 Saito et al. 1996; Momose et al. 1996).  
In this section, we discuss the physical structures
 of the disklike envelopes obtained by our calculations 
 and make a comparison between our numerical results
 and observations.

\begin{figure}
\epsscale{1.0}
\plotone{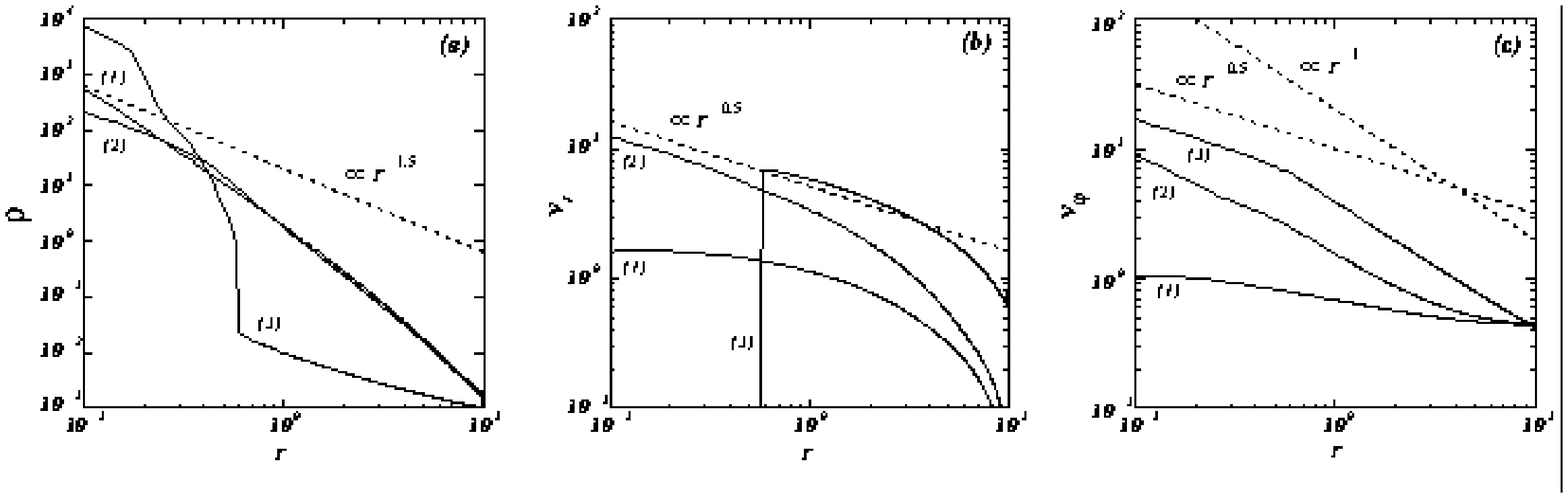}
\caption{
Radial profiles of (a) density, 
 (b) radial velocity, and (c) azimuthal velocity
  in the equatorial plane for model E.
They are plotted at the stages at which 
(1) $t-t_{\rm core}=0.00t_{\rm u}$ 
 (the epoch of core formation), (2) 1.56
 ($M_0\simeq 0.4 M_{\rm t}$), and (3) 13.43
 ($M_0\simeq 0.85 M_{\rm t}$), 
 where $t-t_{\rm core}$ is the time measured from the epoch of core
 formation and $M_{\rm 0}$ is the mass of the central point-like object.  
If we adopt $\rho _0=10^6$cm$^{-3}$ and $c_s=0.2$km s$^{-1}$, 
 then the units of time and length are given as
 $t_{\rm u}=(2\pi G\rho_ 0)^{-0.5}=2.5\times 10^4$ yr and 
 $r_{\rm u}=c_s t_{\rm u}=1.1\times 10^3$ AU, respectively.
The total cloud mass is equal to
 $2.1\times 10^2m_{\rm u}=1.6M_\odot$.
}
\label{fig:profiles}
\end{figure}

Figure \ref{fig:profiles} shows the radial profiles of
 $\rho$, $v_r$, and $v_\varphi$
 in the equatorial plane during the accretion phase.
Just after core formation, the density and velocity
 profiles are well approximated by $\rho \propto r^{-2}$, 
 $v_r \sim {\rm const.}$, and $v_\varphi \sim {\rm const.}$
 near the center.
This indicates that the central high-density region converges
 well onto a similarity solution by the epoch of core formation
 (Matsumoto et al. 1997).
As the collapse progresses, both the infalling disk and 
 the rotationally-supported disk grow in size.
In the infalling disk, the density, radial velocity,
 and  azimuthal velocity have the radial distributions of
 $\rho \propto r^{-1.5}$, $v_r \propto r^{-0.5}$, and
 $v_\varphi \propto r^{-1}$, respectively.
The radial velocity is reproduced by a free-fall velocity
 of $\sqrt{2GM _0/r}$.
The azimuthal velocity means that the specific angular 
 momentum is almost constant in the infalling envelope.
Such dependence seems to be qualitatively similar to 
 that of a similarity solution of an infalling rotating disk
 (Saigo \& Hanawa 1998; see also Terebey, Shu, Cassen 1984).
In the rotationally-supported disk, 
 the radial and azimuthal velocities are well approximated by
 $v_r\sim 0$ and $v_\varphi \propto r^{-0.5}$, respectively.
At the boundary between the infalling disk and 
 the rotationally-supported disk, a shock wave is formed
 and propagates outwardly.

Momose et al. (1996) have examined the physical structure of 
 the infalling disk around the L1551 IRS5.
From a comparison between 
 the observed postion-velocity diagram and their simple
 analytic model, they suggested that the velocity field 
 in the disklike envelope is explained in terms of infall with 
 slow rotation.
Their simple model indicates 
 that the infall and azimuthal velocities have 
 radial dependences of 
 $v_r \propto r^{-0.5}$ with 0.5 km s$^{-1}$ at $r=700$ AU
 and $v_\varphi \propto r^{-1}$ with 0.24 km s$^{-1}$
 at $r=700$ AU, respectively.
If we take $\rho _0 \sim 10^6$ cm$^{-3}$ and $T=10$ K, 
 then the observed velocity field around the L1551 IRS5
 is consistent with that of the model
 presented in \S \ref{subsec:rotation}
 at the stage at which
 $t-t_{\rm core}\sim 1.5(2\pi G\rho _0)^{-0.5}\simeq 3.8\times 10^4$ yr, 
 where $t-t_{\rm core}$ denotes the time measured from the epoch of 
 core formation.
The mass infall rate is evaluated as $\dot{M}\sim 1\times 10^{-5}
M_\odot$ yr$^{-1}$.
The central mass (a point mass plus its 
 surrounding rotationally-supported disk)
 is then estimated as 0.6 $M_\odot$.

The detailed comparisons between the numerical results
 and observations will reveal the formation and
 evolution processes of the infalling envelopes 
 as well as the young stars.

\acknowledgements

We are grateful to T. Hanawa, N. Hirano, M. Momose, Y. Nakamura, N. Ohashi, 
 and K. Tomisaka for useful discussions. 
We also thank an anonymous referee for valuable comments which 
 improve the paper.
Numerical computations were carried out 
on VPP300/16R at the Astronomical Data Analysis Center
 of the National Astronomical Observatory, Japan.
This work was financially supported in part by the Grant-in-Aid 
 for Scientific Research on Priority Areas 
 of the Ministry of Education, Science, Sports and Culture
 (10147205, 11134203).

\begin{deluxetable}{lccccc}
\tablecolumns{5}
\tablewidth{.55\columnwidth}
\tablecaption{MODEL PARAMETERS}
\tablehead{
\colhead{Model} & \colhead{$\beta$} & \colhead{$\lambda _z$}
 & $M_{\rm t}$ & \colhead{$\Delta r=\Delta z$}  & \colhead{$N_r\times N_z$}} 
\startdata
A  & 0    & 10.4 & $1.25\times 10^2$ & $1.01\times 10^{-2}$
 & $1024\times 512$ \\
B  & 0    & 15.5 & $1.93\times 10^2$ & $1.52\times 10^{-2}$
 & $1024\times 512$ \\
C  & 0    & 23.3 & $2.91\times 10^2$ & $2.27\times 10^{-2}$
 & $1024\times 512$ \\
D  & 0.01 & 15.5 & $1.93\times 10^2$ & $1.52\times 10^{-2}$
 & $1024\times 512$ \\
E  & 0.05 & 15.5 & $2.07\times 10^2$ & $1.52\times 10^{-2}$
 & $1024\times 512$ \\
F  & 0.10 & 15.5 & $2.10\times 10^2$ & $1.52\times 10^{-2}$
 & $1024\times 512$ \\
\enddata
\tablecomments{$\beta=$ the ratio of the centrifugal force
 to the pressure force; 
$\lambda _z=$ the longitudinal wavelength; 
$M_{\rm t}=$ total cloud mass; 
$\Delta r$ and $\Delta z=$ the grid spacings
 in the $r$- and $z$-directions;
$N_r$ and $N_z=$ the number of grid points 
 in the $r$- and $z$-directions.}
\label{tab:model}
\end{deluxetable}


\end{document}